\documentclass{aa}

\newcommand{\msun}{\,\mathrm{M}_\odot}
\newcommand{\rsun}{\,\mathrm{R}_\odot}
\newcommand{\kms}{\,\mathrm{km}\,\mathrm{s}^{-1}}

\newcommand{\days}{\,\mathrm{days}}

\newcommand{\au}{\,\mathrm{AU}}

\newcommand{\Myr}{\,\mathrm{Myr}}
\newcommand{\Gyr}{\,\mathrm{Gyr}}
\newcommand{\yr}{\,\mathrm{yr}}
\newcommand{\cs}{\,c_\mathrm{s}}

\def\lesssim{\mathrel{\hbox{\rlap{\hbox{\lower3pt\hbox{$\sim$}}}\hbox{\raise2pt\hbox{$<$}}}}}
\def\lesseq{\mathrel{\hbox{\rlap{\hbox{\lower3pt\hbox{$-$}}}\hbox{\raise2pt\hbox{$<$}}}}}
\def\gtrsim{\mathrel{\hbox{\rlap{\hbox{\lower3pt\hbox{$\sim$}}}\hbox{\raise2pt\hbox{$>$}}}}}
\def\gtreq{\mathrel{\hbox{\rlap{\hbox{\lower3pt\hbox{$-$}}}\hbox{\raise2pt\hbox{$>$}}}}}

\usepackage{xcolor}
\usepackage{graphicx}
\usepackage{subcaption}
\usepackage{txfonts}
\usepackage{natbib,twoopt}
\usepackage[breaklinks=true]{hyperref} 
\bibpunct{(}{)}{;}{a}{}{,}
\usepackage{cleveref}
\usepackage{todonotes}
\usepackage{diagbox}
\usepackage{esvect}
\makeatletter
  \newcommandtwoopt{\citeads}[3][][]{\href{http://adsabs.harvard.edu/abs/#3}%
    {\def\hyper@linkstart##1##2{}%
     \let\hyper@linkend\@empty\citealp[#1][#2]{#3}}}
  \newcommandtwoopt{\citepads}[3][][]{\href{http://adsabs.harvard.edu/abs/#3}%
    {\def\hyper@linkstart##1##2{}%
     \let\hyper@linkend\@empty\citep[#1][#2]{#3}}}
  \newcommandtwoopt{\citetads}[3][][]{\href{http://adsabs.harvard.edu/abs/#3}%
    {\def\hyper@linkstart##1##2{}%
     \let\hyper@linkend\@empty\citet[#1][#2]{#3}}}
  \newcommandtwoopt{\citeyearads}[3][][]%
    {\href{http://adsabs.harvard.edu/abs/#3}
    {\def\hyper@linkstart##1##2{}%
     \let\hyper@linkend\@empty\citeyear[#1][#2]{#3}}}
\makeatother
\begin{document}

   \title{Protostellar spin-up and fast rotator formation through binary star formation}
   \titlerunning{fast rotator formation pathway}


   \author{Rajika L.~Kuruwita
          \inst{1},
          Christoph Federrath\inst{2,3}
          \and
          Marina Kounkel\inst{4}
          }

   \institute{Heidelberg Institute for Theoretical Studies, Schlo{\ss}-Wolfsbrunnenweg 35, 69118 Heidelberg, Germany\\
              \email{rajika.kuruwita@h-its.org}
            \and
             Research School of Astronomy and Astrophysics, Australian National University, Canberra, ACT 2611, Australia
             \and
             Australian Research Council Centre of Excellence in All Sky Astrophysics (ASTRO3D), Canberra, ACT 2611, Australia
             \and
             Department of Physics, University of North Florida, 1 UNF Dr, Jacksonville, FL 32224, USA
             }

   \date{Received April 22, 2024; accepted August 24, 2024}

 
  \abstract
   {Many fast rotating stars (rotation periods of $<2\days$) are found to be unresolved binaries with separations of tens of au. This correlation between fast rotators and binarity leads to the question of whether the formation of binary stars inherently produces fast rotators.}
   {Our goal is to understand the spin evolution of protostars and whether the formation of companions plays a role in spinning up stars.}
    {We use magneto-hydrodynamical simulations to study the formation of multiple star systems from turbulent and non-turbulent protostellar cores. We track the angular momentum accreted by individual star and inner disc systems by using a sink (star) particle technique. We run a resolution study to extrapolate protostellar properties.}
   {We find in all simulations that the primary star can experience a spin-up event correlated with the formation of companions, i.e., fragmentation into binaries or higher-order systems. The primary star can spin up by up to 84\% of its pre-fragmentation angular momentum and by up to 18\% of its pre-fragmentation mass-specific angular momentum. The mechanism for the spin-up is gravitational disc instabilities in the circumstellar disc around the primary star, leading to the accretion of material with high specific angular momentum. The simulations that experience the strongest disc instabilities fragment to form companions. Simulations with weaker spin-up events experience disc instabilities triggered by a companion flyby, and the disc instability in these cases typically does not produce further fragments, i.e., they remain binary systems.}
   {The primary star in multiple star systems may end up with a higher spin than single stars. This is because gravitational instabilities in the circumstellar disc around the primary star can trigger a spin-up event. In the strongest spin-up events, the instability is likely to cause disc fragmentation and the formation of companions. This spin-up mechanism, coupled with shorter disc lifetimes due to truncated circumstellar discs (and thus short spin-down times), may help produce fast rotators.}


   \maketitle
%

\section{Introduction}

During the early star formation process (Class 0/1), a protostar will accrete mass from its surrounding circumstellar disc. With the inflow of the material onto the protostar, it will also accrete angular momentum, and the resulting rotational velocity can increase by any amount up to the break-up velocity, which is defined as the Keplerian velocity at the stellar surface. The initial spin of the protostar is therefore set by the angular momentum of the accreted gas, as well as the interactions between the star and circumstellar disc. 

While protostars spin up due to both contraction and accretion, they are also expected to spin down later in their protostellar evolution, due to disc-locking with the circumstellar disc \citep{konigl_disk_1991}. Observations of stellar spins in various star-forming regions $<4\Myr$ (ages when circumstellar discs are still present) find that stars do spin down slowly, however, there is a significant spread in rotation rates at all ages \citep{rebull_stellar_2004}. The difference in the angular velocity between the protostar and the inner disc leads to angular momentum being removed from the protostar, slowing down its rotation. The presence of a disc is, therefore, crucial to the spin-down process.

Models that explore the evolution of stellar rotation over a period of up to $1\Gyr$ after formation find that shorter disc lifetimes can explain the evolution of fast rotators \citep{eggenberger_impact_2012, gallet_improved_2013, gallet_improved_2015}, as a shorter period of disc-locking will remove less angular momentum. Observations also support the relation between disc lifetime and rotation rates, with disc-bearing pre-main sequence (PMS) stars consistently having longer rotation periods ($3-10\days$) than disc-less PMS stars ($1-7\days$), \citep{edwards_angular_1993, cieza_testing_2007, affer_rotation_2013, kounkel_measurement_2023}. The observations imply that fast disc dispersal halts the spin-down phase.

An origin of fast disc dispersal can be the presence of a companion, which truncates the circumstellar disc \citep{kraus_impact_2016, kuruwita_multiplicity_2018}, thus leading to short disc lifetimes and spin-down times. This potential relationship between binary and fast rotator formation has been discussed in the literature, with \citet{stauffer_angular_1994} suggesting binary stars with orbital periods of ten to tens of years are likely to be fast rotators due to less or no disc-regulated spin evolution due to disc truncation by a companion.

Early observational surveys investigating a relationship between the presence of companions and spin evolution for binaries with separations from $10-1000$ astronomical units ($\au$) found no significant relationship \citep{bouvier_pleiades_1997}, however, these observations look at the $100\Myr$ Pleiades, which is very evolved. More recent observations of younger stars find that many fast rotators (period $<2\days$) are found to be in binary star systems with an unseen companion at separations $\sim10$s of au, and this can be observed both in evolved systems, as well as those that have ages of $<1$ Myr. \citep{daher_stellar_2022, kounkel_measurement_2023}

While shortened disc lifetimes are helpful for fast rotator formation, this explanation is insufficient as a sole factor for driving the excess in angular momentum in these stars, as binaries that can retain their discs for $\sim$5 Myr are still somehow able to eventually spin up. Nor do these models explain the initial rotation period distribution of the stars that have been modelled. What sets the initial spin of a star is still an open question. 

Previous simulations of binary star formation have found that the amount of angular momentum that is transported via outflows is less efficient in young binary star systems (with separations of a few tens of au) which can lead to more angular momentum remaining in the star-disc system \citep{kuruwita_binary_2017}, giving the first hint of the role of multiplicity on the evolution of angular momentum in young stars.

In this paper, we investigate directly how protostars obtain their initial spin, and whether fragmentation plays a role in spinning up stars. In \Cref{sec:method} we describe the simulations used, and the sink particle model that allows us to track the spin of the protostars. In \Cref{sec:results}, we present the results of our simulations and determine a mechanism for spinning up a protostar. In \Cref{sec:discussion} we discuss the plausibility of our mechanisms based on comparison with observations. In \Cref{sec:caveats} we discuss the limitations of the work and the impact of missing physics. In \Cref{sec:conclusion} we summarise the conclusions of this study.

\section{Method}
\label{sec:method}

We use the adaptive mesh refinement (AMR) code \texttt{FLASH} \citep{fryxell_flash:_2000, dubey_challenges_2008} to integrate the compressible ideal magneto-hydrodynamic (MHD) equations. Here we use the HLL3R Riemann solver for ideal MHD \citep{waagan_robust_2011}. The gravitational interactions of the gas are calculated using a tree-based Poisson solver \citep{wunsch_tree-based_2018}.

\subsection{Equation of state}

We use a piece-wise polytropic equation of state, given by
\begin{equation}
P_\mathrm{th} = K\rho^\Gamma,
\label{eqn:eos}
\end{equation}
which describes the relationship between the thermal gas pressure ($P_\mathrm{th}$) and density ($\rho$), where $K$ is the polytropic coefficient and $\Gamma$ is the polytropic index. The coefficient $K$ is given by the isothermal sound speed squared. In our simulations, the sound speed is initially set to $c_\mathrm{s} = 2 \times 10^4\,\mathrm{cm}\,\mathrm{s}^{-1}$ for a gas temperature of $\sim11\,\mathrm{K}$ with mean molecular weight of $2.3\,m_\mathrm{H}$, where $m_\mathrm{H}$ is the mass of a hydrogen atom. $K$ is then subsequently computed, such that $P$ is a continuous function of $\rho$. For our simulations $\Gamma$ is defined as
\begin{equation}
\Gamma=\begin{cases}
1.0  \text{ for \,\,\,\,\,\,\,\,\,\,\, $\rho \leq \rho_1 \equiv 2.50 \times 10^{-16}\,\mathrm{g}\,\mathrm{cm}^{-3}$},\\
1.1  \text{ for $\rho_1 < \rho \leq \rho_2 \equiv 3.84 \times 10^{-13}\,\mathrm{g}\,\mathrm{cm}^{-3}$},\\
1.4  \text{ for $\rho_2 < \rho \leq \rho_3 \equiv 3.84 \times 10^{-8}\,\mathrm{g}\,\mathrm{cm}^{-3}$},\\
1.1  \text{ for $\rho_3 < \rho \leq \rho_4 \equiv 3.84 \times 10^{-3}\,\mathrm{g}\,\mathrm{cm}^{-3}$},\\
5/3 \text{ for \,\,\,\,\,\,\,\,\,\,\,\,$\rho > \rho_4$}.
\end{cases}
\label{eqn:gamma}
\end{equation}

The values of $\Gamma$ were approximated based on radiation-hydrodynamical simulations of molecular-core collapse by \citet{masunaga_radiation_2000}. These values describe the gas behaviour during the initial isothermal collapse of the molecular core, adiabatic heating of the first core, the H$_2$ dissociation during the second collapse into the second core, and the return to adiabatic heating.

\subsection{Adaptive mesh refinement (AMR)}

The Jeans length must be resolved with at least 4~grid cells to avoid artificial fragmentation \citep{truelove_jeans_1997}. However, it is recommended that the Jeans length be resolved with at least 30~grid cells to resolve the twisting and amplification of the magnetic field, and the turbulent motions on the Jeans scale \citep{federrath_implementing_2011}. In our simulations, the Jeans length is resolved with at least $32$~grid cells. This Jeans refinement criterion applies on all AMR levels except the highest level of the AMR hierarchy. On the maximum level of refinement, the formation of sink (star) particles may be triggered, which is discussed next.

\subsection{Sink Particles}
\label{ssec:sinkparticles}

\subsubsection{Sink particle formation}

In our simulations, the formation of a protostar is signalled by the formation of a sink particle \citep{federrath_modeling_2010, federrath_modeling_2014}. If a cell exceeds the density threshold, derived from the Jeans length, given by
\begin{equation}
\rho_\mathrm{sink}= \frac{\pi\cs^2}{4Gr_\mathrm{sink}^2},
\label{eqn:densitythreshold}
\end{equation}
it may collapse into a point-mass sink particle (subject to additional checks; see below). 

For a gas volume with a central cell that exceeds $\rho_\mathrm{sink}$ to form a sink particle, the gas volume must also meet the following criteria described in \cite{federrath_modeling_2010}:
\begin{enumerate}
	\item Be on the highest level of AMR.
	\item Not be within the accretion radius ($r_\mathrm{sink}$) of an existing sink particle (c.f.~\Cref{ssec:sink_accretion}).
	\item The velocity field must be converging from all directions ($v_\mathrm{r} < 0$).
	\item Have a central gravitational potential minimum.
	\item Be bound ($|E_\mathrm{grav}| > E_\mathrm{th} + E_\mathrm{kin} + E_\mathrm{mag}$).
	\item Be Jeans-unstable.
\end{enumerate}

If these criteria are met for a volume with radius $r_\mathrm{sink}=2.5 \Delta x$, where $\Delta x$ is the cell length on the highest AMR level, centred on the cell exceeding $\rho_\mathrm{sink}$, then a sink particle is created.

Once a sink particle is formed, a second-order leapfrog integrator is used to update the positions of sink particles using a time step based on the velocity and acceleration of the sink particle. A sub-cycling method is implemented to prevent artificial precession of the sink particles in eccentric orbits. Details on the sub-cycling method can be found in \citet{federrath_modeling_2010}, highlighting that this method is necessary to accurately calculate sink particle orbits. 

\subsubsection{Sink particle accretion model}
\label{ssec:sink_accretion}

As the simulations evolve, sink particles that form can accrete mass and momentum. A sink particle can accrete mass if a cell exceeds the density threshold $\rho_\mathrm{sink}$ while within the accretion radius $r_\mathrm{sink}$ of a sink particle. If the mass is bound and collapses towards the sink particle, it will be accreted onto the sink particle, such that mass, momentum and angular momentum are conserved \citep{federrath_modeling_2010,federrath_modeling_2014}. It is with this model that we can track how much angular momentum the protostars will accrete.

The gravitational binding energy of a parcel of gas alone does not predict whether the gas will be accreted onto the star. Gas is accreted onto a sink particle only if it is bound and collapsing towards the particle, i.e., $v_\mathrm{r} < 0$. The fraction of mass $\Delta m_i$ to be accreted from cell $i$ with density $\rho_i$ and volume $V_i$ is $\Delta m_i = (\rho_i - \rho_\mathrm{sink})V_i$.

Within the control volume $(4\pi/3)r^{3}_\mathrm{sink}$ of a sink particle, we calculate the mass, centre of mass (c.o.m.), momentum, and angular momentum of the material to be accreted:
\begin{equation}
  \begin{gathered}
    \mathrm{mass}: \mathrm{M}_\mathrm{acc} = \sum_\mathrm{i}\Delta m_\mathrm{i} \\
    \mathrm{c.o.m.}: \mathrm{M}_\mathrm{acc}\mathbf{R}_\mathrm{acc} = \sum_\mathrm{i}\Delta m_\mathrm{i} \mathbf{r}_\mathrm{i} \\
    \mathrm{momentum}: \mathrm{M}_\mathrm{acc}\mathbf{V}_\mathrm{acc} = \sum_\mathrm{i} \Delta m_\mathrm{i} \mathbf{v}_\mathrm{i} \\
    \mathrm{ang.mom.}: \mathbf{L}_\mathrm{acc} = \sum_\mathrm{i} \Delta m_\mathrm{i} \mathbf{r}_\mathrm{i} \times \mathbf{v}_\mathrm{i}
  \end{gathered}
  \label{eqn:accretion_quantity}
\end{equation}

\begin{figure*}
    \centerline{\includegraphics[width=\linewidth]{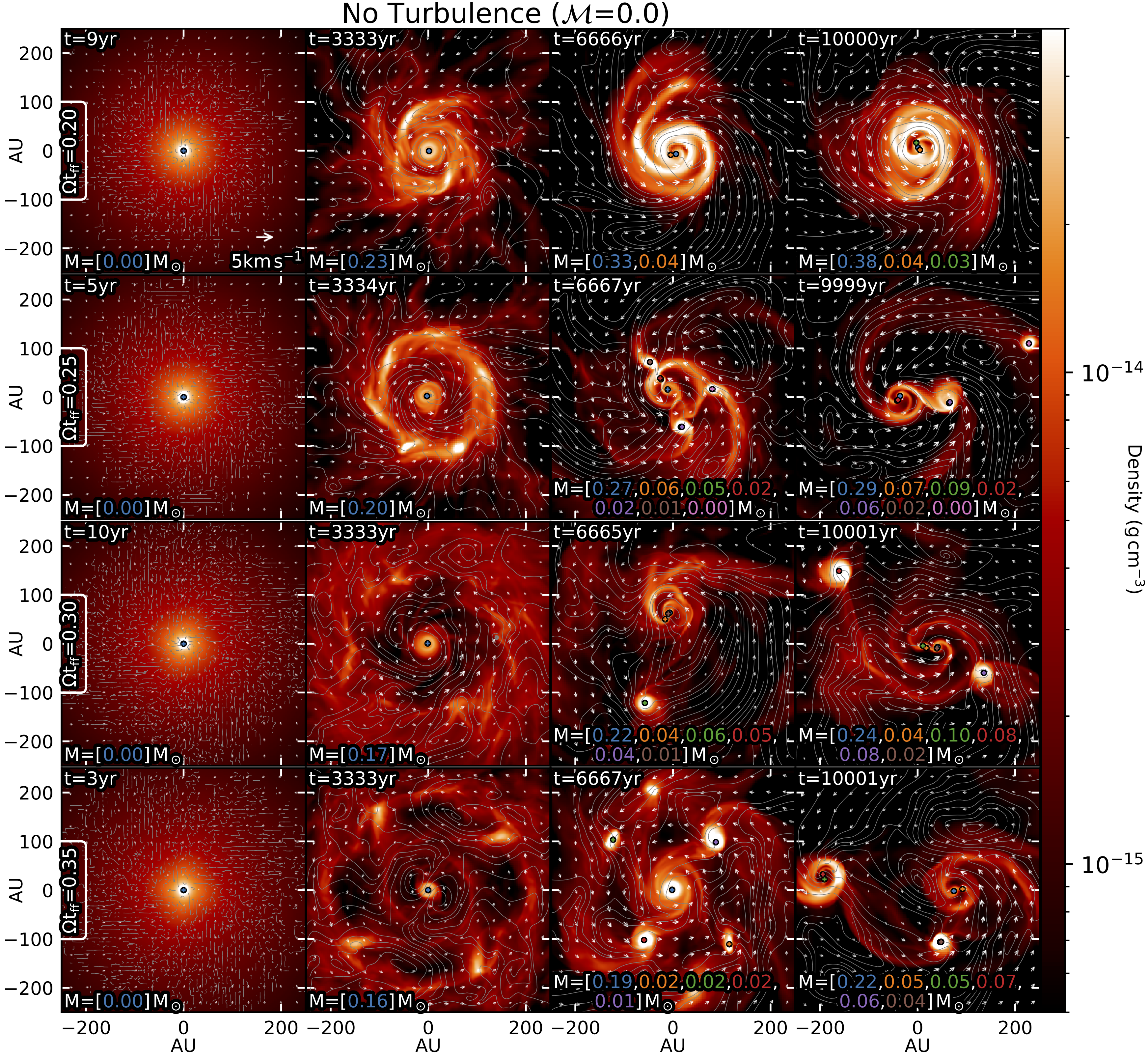}}
    \caption{Projections of the density over $200\au$ thick slices at different times, for the non-turbulent simulations. The rows from top to bottom show simulations from low to high initial spin ($\Omega t_\mathrm{ff}=0.20$ to $0.35$). The time since the formation of the primary star is annotated in the top left of each panel. The columns are the simulations at time $0, 3333, 6666$ and $10\,000\yr$ after the first proto-star forms. The sink particles are annotated with coloured crosses, and their masses are annotated at the bottom of each panel. The vector field shows the integrated velocity field, and the streamlines show the magnetic field.}
    \label{fig:simulation_projections_M0}
\end{figure*}

The mass $\Delta m_\mathrm{i}$ is then removed from each cell and added to the sink particle quantities are updated with the accreted amount. We refer the reader to \citet{federrath_modeling_2014} for details on the sink particle module.

Throughout the simulations, we can track how the spin of the sink particles evolves, and we use this property to gain insight into the origin of fast rotators.

\subsection{Simulation setup}
\label{sec:simulation_setup}

Our simulations are initialised in a three-dimensional computational domain of $L_\mathrm{box}=1.2\times10^{17}\,\mathrm{cm}$ ($\sim$8000$\au$) along each side of the Cartesian domain. We use 12~levels of refinement ($L_\mathrm{ref}$) of the AMR grid, resulting in a minimum cell size of $1.95\au$ when fully refined. At this resolution, the accretion radius of the sink particles is $4.9\au$. A resolution study was conducted on one of the simulations (c.f.~\Cref{ssec:resolution_study}) where we determined how the sink particle properties depend on resolution, allowing us to extrapolate to the properties of a protostar with radius $r=2\rsun$.

\begin{figure*} \centerline{\includegraphics[width=\linewidth]{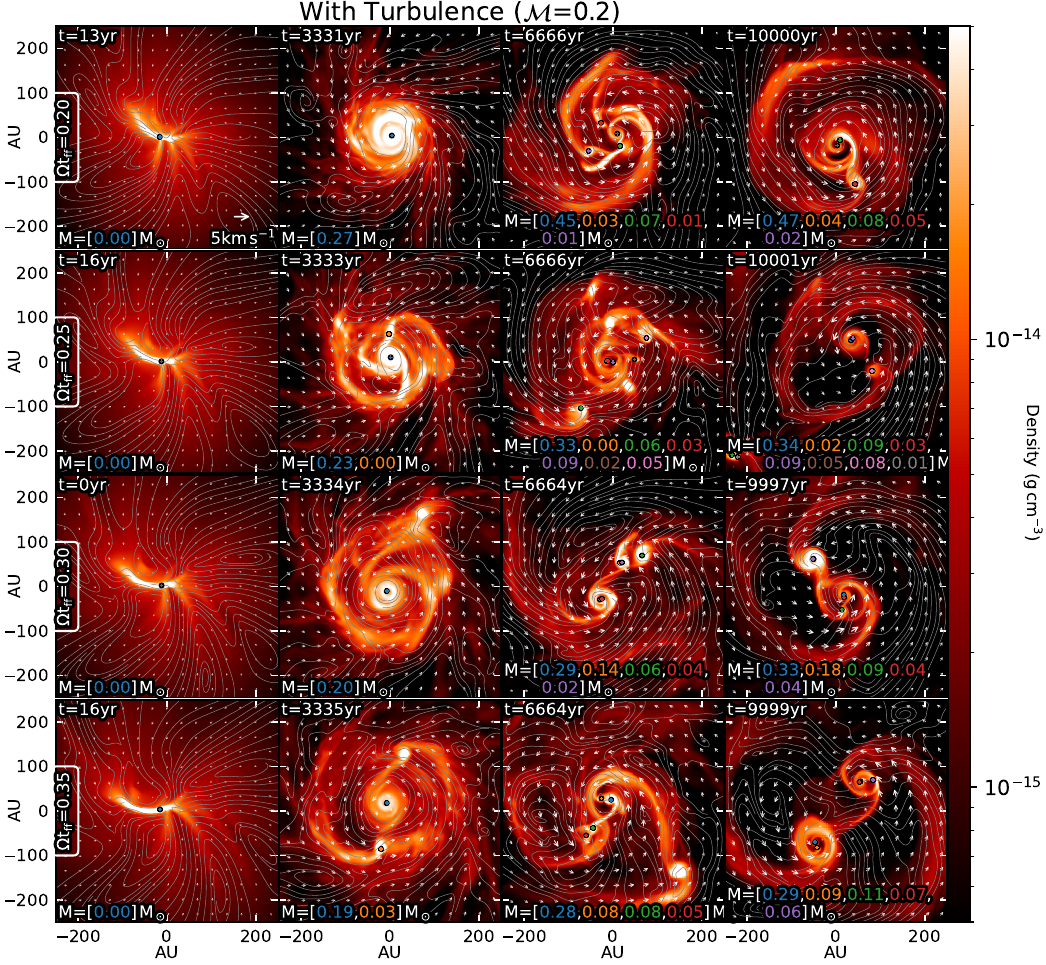}}
    \caption{Same as \Cref{fig:simulation_projections_M0}, but for the simulations with turbulence (Mach number $\mathcal{M}=0.2$).}
    \label{fig:simulation_projections_M2}
\end{figure*}

Our simulations begin with a spherical cloud of mass $1\msun$, and radius $\sim$3300$\au$ placed in the centre of the simulation domain, which is initially given solid body rotation. We run a parameter study with varying initial cloud rotation and turbulence levels. The rotation is defined as values of $\Omega t_\mathrm{ff}$ (see \cite{banerjee_outflows_2006} and \cite{machida_high-_2008}) from $0.20$ to $0.35$, where $\Omega$ is the angular frequency, and $t_\mathrm{ff}$ is the freefall time of the cloud. A higher $\Omega t_\mathrm{ff}$ value indicates a higher initial rotation rate. An initial turbulent velocity field is imposed on top of the solid body rotation. We perform runs without turbulence and with a typical level of turbulence of Mach number $0.2$, which corresponds to a velocity dispersion of $0.04\kms$. For details on the implementation of turbulence, we refer the reader to \citet{federrath_comparing_2010} and \citet{kuruwita_role_2019}.

An initially uniform magnetic field of $100\,\mu\mathrm{G}$ is threaded through the cloud in the \emph{z}-direction. This gives a mass-to-flux ratio of $(M/\Phi)/(M/\Phi)_\mathrm{\mathrm{crit}}=5.2$ where the critical mass-to-flux ratio is $487\,\mathrm{g}\,\mathrm{cm}^{-2}\,\mathrm{G}^{-1}$ as defined in \citet{mouschovias_note_1976}. This value makes the protostellar core magnetically super-critical, such that it will collapse under its own gravity.

To prevent the cloud from expanding rapidly due to internal gas pressure, the spherical cloud is initialised in pressure equilibrium with the surrounding gas material. This is achieved by giving the surrounding material a gas density of $\rho_0/100$ with a temperature 100~times higher than the cloud temperature.

\section{Results}
\label{sec:results}

\begin{table}
	\centering
	\begin{tabular}{|l|c|c|c|c|}
		\hline
		  \diagbox[width=7em]{$\mathcal{M}$(km/s)}{$\Omega t_\mathrm{ff}$}  & 0.20 & 0.25 & 0.30 & 0.35\\
    \hline
          0.0 (0.0) & 3 & 7 & 6 & 6\\
          0.2 (0.02)& 5 & 8 & 5 & 5\\
		\hline
	\end{tabular}
	\caption{Number of stars formed by $10\,000\,\yr$ after primary star formation in each simulation.}
	\label{tab:N_stars}
\end{table}

\begin{figure}
    \centerline{\includegraphics[width=\linewidth]{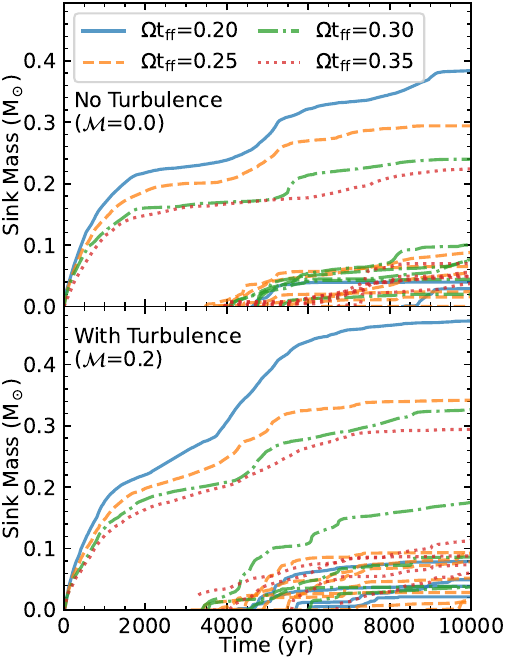}}
    \caption{The mass of the sink particles since the formation of the primary star for simulations without turbulence (\emph{top}) and with turbulence (\emph{bottom}). The initial spin of the simulations is indicated in the legend. Each line shows the mass of an individual sink particle.}
    \label{fig:sink_mass_evolution}
\end{figure}

\begin{figure}
    \centerline{\includegraphics[width=\linewidth]{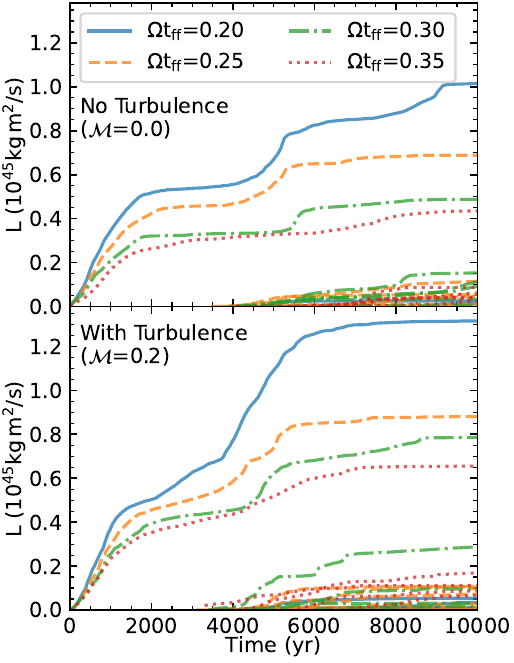}}
    \caption{Same as \Cref{fig:sink_mass_evolution}, but for the total angular momentum accreted by the sink particles.}
    \label{fig:sink_spin}
\end{figure}

\begin{figure}
    \centerline{\includegraphics[width=\linewidth]{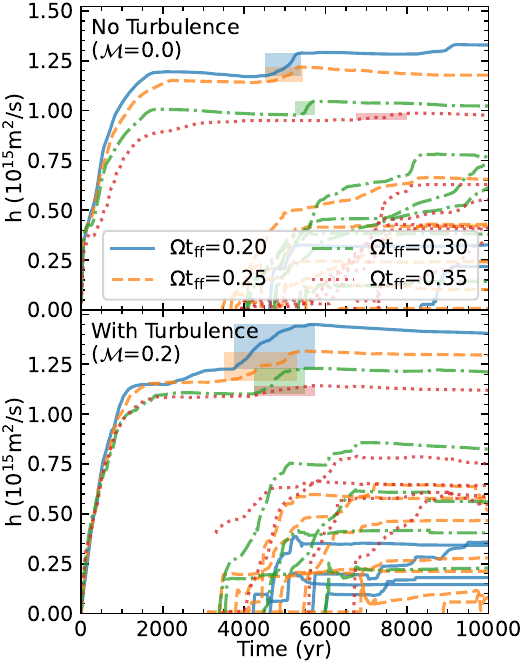}}
    \caption{Same as \Cref{fig:sink_mass_evolution}, but for the mass-specific angular momentum accreted by the sink particles. The highlighted regions indicate the first spin-up event after the formation of companions.}
    \label{fig:sink_spin_spec}
\end{figure}

\begin{figure}
    \centerline{\includegraphics[width=\linewidth]{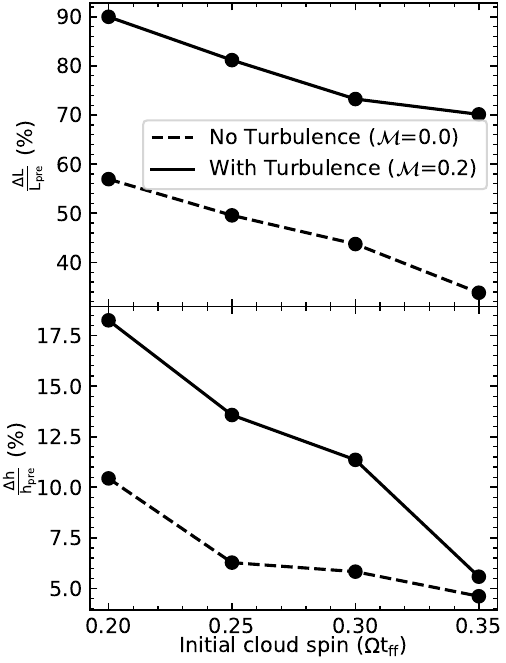}}
    \caption{Measured spin-up calculated relative to the angular momentum ($\Delta L$; \emph{top}) and specific angular momentum ($\Delta h$; \emph{bottom}) of the primary star at $500\yr$ before the formation of the secondary companion, and the peak momentum in the highlighted regions in \Cref{fig:sink_spin_spec}.}
    \label{fig:spin_up}
\end{figure}

Our simulations evolve for at most $10\,000\yr$ after the first sink particle (hereafter referred to as the primary star) is formed. The entire suite of simulations used approximately a million compute core hours. The number of sink particles produced by the end of each simulation is summarised in \Cref{tab:N_stars}. Projection plots of the density at various times are shown in \Cref{fig:simulation_projections_M0} and \Cref{fig:simulation_projections_M2} for the non-turbulent and turbulent cases, respectively. In \Cref{fig:simulation_projections_M0}, where the primary star first forms, the density distribution of the gas is rotationally symmetric, and smooth. For the simulation with the lowest initial spin in the top row, we see that a single large circumstellar disc forms around the primary, and companions then form within this disc. For the higher initial spin simulations, a dense ring quickly forms, which then fragments into multiple stars with initial separations of $\sim$100$\au$. After fragmentation, there are complex dynamics that cluster the stars into different groups, forming circum-binary and -trinary discs. For the turbulent simulations shown in \Cref{fig:simulation_projections_M2}, we see that when the primary star first forms, the initial turbulent velocity field creates non-axisymmetric density perturbations. The lowest initial spin simulations form a large circumstellar disc, which then fragments, similar to the non-turbulent case. For the higher initial spin simulations, an extended circumstellar disc forms, which is in contrast to the ring formed in the non-turbulent simulations. The initial turbulence velocity field smears out what would be a ring into an extended low-density disc. In the turbulent simulations, which form many sink particles, the dynamics seen in the high initial spin non-turbulent simulations are also present here.

The mass evolution of the sink particles in all simulations is shown in \Cref{fig:sink_mass_evolution}. We find that the primary star shows an initial rapid growth in mass over the first $\sim$1500$\yr$ before plateauing. In most simulations, the primary star also experiences an increased growth period later on (around $4000-5000\yr$), which coincides with the formation of companions.

Before the second mass growth phase, the initial turbulence of the simulation does not directly affect the primary star mass anymore. However, after the second mass growth phase, we find that the amount of turbulence has some effect on the primary star, with the final primary mass being larger in the simulations with initial turbulence. The initial spin of the proto-stellar cloud shows a significant difference in the mass evolution of the primary stars, with the lower spin simulations resulting in more massive primary stars. For the $\mathcal{M}$=0.0, the final primary mass of the $\Omega t_\mathrm{ff}$=0.20 simulation is approximately 70\% larger than the final primary mass of the $\Omega t_\mathrm{ff}$=0.35 simulation. In the $\mathcal{M}$=0.2, the primary in the lowest spin simulation is approximately 62\% larger than the primary in the highest spin simulation. This may be because the surrounding gas in the higher initial spin simulations must lose significantly more angular momentum to fall within the accretion radius of the primary star.

\subsection{Protostellar spin}

While the sink particles accrete mass throughout the simulations, they also accrete angular momentum. In \Cref{fig:sink_spin} the total angular momentum of the individual sink particles is shown. In all simulations, we see that the primary star experiences an initial growth in angular momentum and then plateaus. In most of the simulations, we also see a second phase of angular momentum growth, which coincides with the formation of companions. The angular momentum evolution of the sink particles is similar evolution to their mass evolution. This is not surprising, given that angular momentum is directly proportional to the mass. We therefore also present the mass-specific angular momentum ($h$) in \Cref{fig:sink_spin_spec}. In this figure, we see that the specific angular momentum of the primary stars grows significantly over the first 1-2~thousand years, and plateaus. In some cases, we see a steady decrease after the initial mass accretion event. This indicates that after the initial growth, the gas that is being accreted is low in angular momentum, thus reducing the sink particle's specific angular momentum.

In most of the simulations, we see the specific angular momentum of the primary stars increase at approximately the same time as the formation of companions occurs, after $\sim 3000\yr$. This increase indicates that the primary stars are accreting high angular momentum material that then increases the specific angular momentum of the sink particle. From the specific angular momentum evolution, we identify by eye the first significant spin-up event. These spin-up events are highlighted by the shaded region in \Cref{fig:sink_spin_spec}.

We measure the increase in angular momentum and specific angular momentum of the primary star caused by companion formation, using
\begin{equation}
    \frac{\Delta L}{L_\mathrm{pre}} = \frac{L_\mathrm{post}-L_\mathrm{pre}}{L_\mathrm{pre}}, \,\mathrm{and},\, \frac{\Delta h}{h_\mathrm{pre}} = \frac{h_\mathrm{post}-h_\mathrm{pre}}{h_\mathrm{pre}},
    \label{eqn:spin_up_frac}
\end{equation}
where $L_\mathrm{pre}$ and $h_\mathrm{pre}$ are the absolute and specific angular momentum, respectively, of the primary star measured at $500\yr$ before the formation of the first companion. $L_\mathrm{post}$ and $h_\mathrm{post}$ are the maximum absolute and specific momentum, respectively, measured in the primary star, in the highlighted regions of \Cref{fig:sink_spin_spec}. The measured spin-up is shown in \Cref{fig:spin_up}. In this figure, we see that the turbulent simulations consistently produce stronger spin-up events, with an increase of over $70\%$ in the total angular momentum, and high spin-up percentages in the specific angular momentum (up to $18\%$). Across all simulations, we see that there is a trend in that the simulations with low initial spin produce stronger spin-up events than the simulations with higher initial spin. The trend is parallel between the non-turbulent and turbulent simulations, implying that turbulent discs generally lead to higher angular momentum accretion. The trend in specific angular momentum is not parallel between the non-turbulent and turbulent simulations and instead converges at higher initial cloud spin.

\begin{figure*} 
\centerline{\includegraphics[width=\linewidth]{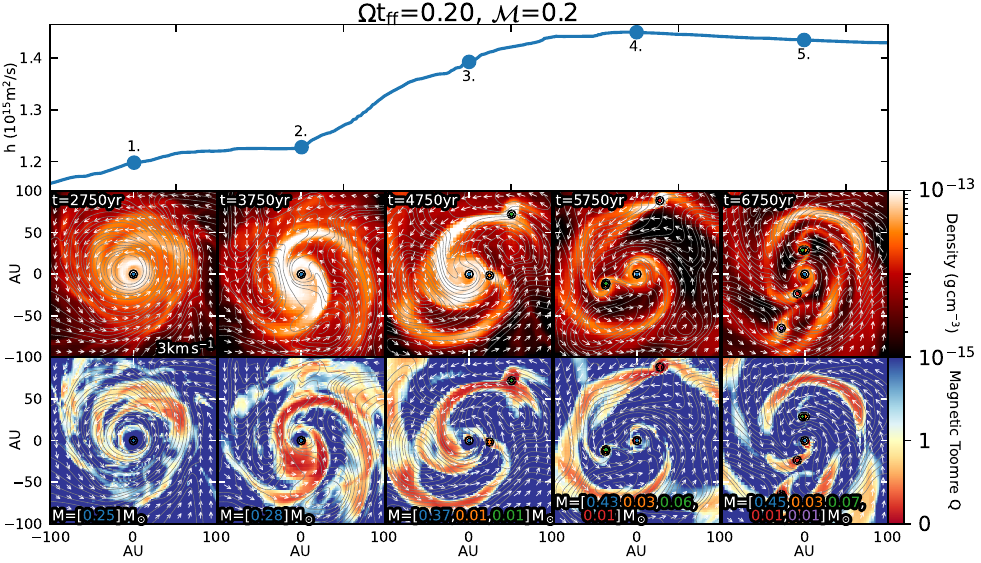}}
    \caption{\emph{top}: the specific angular momentum of the primary around the spin-up event highlighted in \Cref{fig:sink_spin_spec} for the turbulent $\Omega t_\mathrm{ff}=0.20$ simulation. The points on the curve are the times that are shown in the projections below. The second and fourth points are the beginning and end of the spin-up event, and the third point is the middle of the spin-up. The \emph{middle} and \emph{bottom} rows show the density projections (calculated in the same way as \Cref{fig:simulation_projections_M0}) and the magnetic Toomre $Q$, respectively, at the times indicated by the points in the top panel.}
    \label{fig:Toomre_Q_S20M2}
\end{figure*}

The angular momentum accreted onto the sink particles does not directly translate to the angular momentum that would be accreted onto the protostar. The sink particles in the simulations have accretion radii of $4.9\au$ and therefore represent the protostar and inner disc system together. The spin of the protostar is sensitive to the fraction of mass and angular momentum that the star accretes from the inner disc, and what is lost via jets. In the resolution study presented in \Cref{ssec:resolution_study}, we explain that it is not possible to resolve down to the proto-stellar surface while achieving a sufficiently long evolution of these systems. However, our chosen resolution sufficiently some of the disc winds and converges on a consistent sink particle angular momentum evolution before companion formation. 

For the rest of this paper, we assume that the higher angular momentum in the sink particle translates to higher angular momentum in the protostar. If we take a sphere with solid body rotation, the rotational period is given by
\begin{equation}
    P = \frac{4\pi}{5}\frac{r^2}{h},
    \label{eqn:rotation_rate}
\end{equation}
where $r$ is the radius and $h$ is the mass-specific angular momentum of the sphere. For our simulations $4\pi r^2/5=constant$ and with the assumptions we made, the specific angular momentum of the protostar is proportional to the specific angular momentum of the sink particle ($h_\mathrm{\star} \propto h_\mathrm{sink}$), and therefore
\begin{equation}
    P_\star \propto \frac{1}{h_\mathrm{sink}}.
    \label{eqn:Trot_assumption}
\end{equation}

Using this relationship we derive that the post spin-up event period ($P_\mathrm{post}$) can be defined as, $P_\mathrm{post} = P_\mathrm{pre} (h_\mathrm{pre}/h_\mathrm{post})$, where $P_\mathrm{pre}$ is the period of the primary star before the spin-up event. From our simulations we find that a single spin-up event can increase the specific angular momentum by up to 18\%, therefore $P_\mathrm{post} = P_\mathrm{pre} (1/1.18) \sim0.85P_\mathrm{pre}$. Therefore a spin-up event of the magnitude measured in our strongest simulation can reduce the stellar rotation rate by around 15\%.

\subsection{Mechanism for primary star spin up}

We have demonstrated that the primary star in our simulations often experiences a spin-up event after the formation of companions. We aim to understand the mechanisms that trigger a spin-up event by investigating the stability of the disc material near the primary star during a spin-up event. In particular, we look at the simulations that show the strongest and weakest spin-up events, which are the turbulent $\Omega t_\mathrm{ff}=0.20$ and non-turbulent $\Omega t_\mathrm{ff}=0.35$ simulations, respectively.

We determine the stability of material around the primary by calculating the magnetic Toomre $Q$ parameter \citep{forgan_fragmentation_2017}. This is an extension of the classic Toomre $Q$ \citep{toomre_gravitational_1964} to include support from magnetic fields within a disc against collapse. The Toomre $Q$ for a parcel of gas in a near Keplerian disc is calculated as
\begin{equation}
    Q = \frac{c_s\Omega}{\pi G \Sigma},
    \label{eqn:Toomre_Q}
\end{equation}

\begin{figure*} 
\centerline{\includegraphics[width=\linewidth]{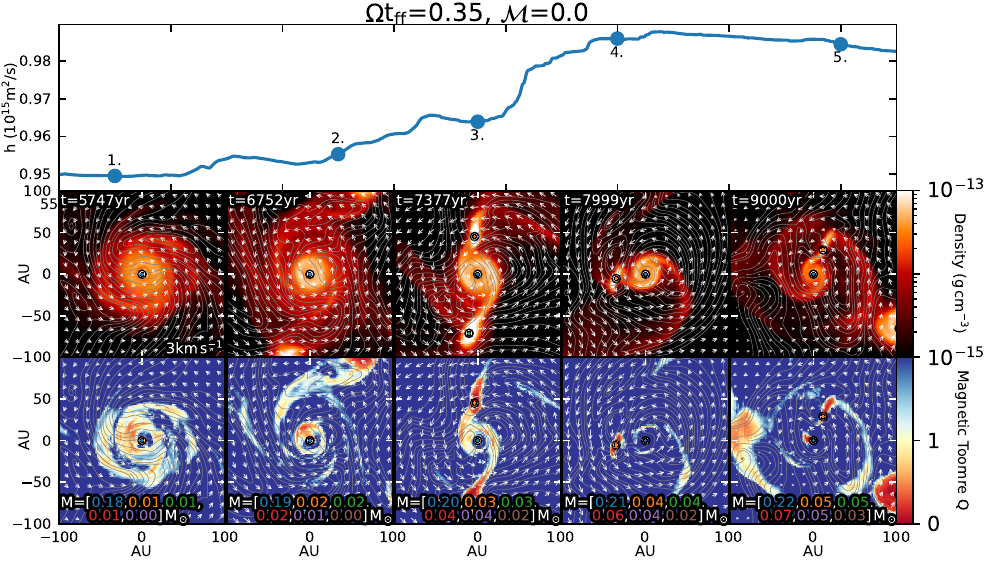}}
    \caption{Same as \Cref{fig:Toomre_Q_S20M2}, but for the non-turbulent $\Omega t_\mathrm{ff}=0.35$ simulation.}
    \label{fig:Toomre_Q_S35M0}
\end{figure*}

where $c_s$ is the sound speed, $\Omega$ is the angular frequency (the tangential velocity divided by the radius, the same used in \Cref{sec:method}), $G$ is the gravitational constant, and $\Sigma$ is the gas surface density of the disc. The surface density in our simulations is derived from projections calculated over $200\au$ thick slices, the same as those shown in \Cref{fig:simulation_projections_M0} and \ref{fig:simulation_projections_M2}. A parcel of gas is considered to be stable if $Q\gg1$, and unstable and prone to collapse if $Q\ll1$. The magnetic Toomre $Q$ is calculated by multiplying the Toomre $Q$ by a scaling factor,

\begin{equation}
    Q_B = Q\sqrt{1 + \beta^{-1}},
    \label{eqn:Mag_Toomre_Q}
\end{equation}

where $\beta$ is the plasma beta of the gas parcel. This $\beta$ factor arises from modifying the sound speed by adding the magnetic pressure to the thermal pressure \citep[e.g.,][]{federrath_star_2012, forgan_fragmentation_2017}.

The resulting magnetic Toomre $Q$ distributions are shown in \Cref{fig:Toomre_Q_S20M2} and \ref{fig:Toomre_Q_S35M0}, for the strongest and weakest spin-up cases, respectively. For the strongest spin-up event (\Cref{fig:Toomre_Q_S20M2}), we see that the spin-up begins before the formation of companions. In the first panel, i.e., $1000\yr$ before the event starts, we see that the primary star hosts a large circumstellar disc of radius $\sim$100$\au$. The inner disc is very stable ($Q\sim10$), but the outer disc is marginally unstable ($Q\sim1$). This instability then grows to form strongly unstable spiral arms, making the inner disc unstable as well in the subsequent evolution. This triggers the spin-up event, funnelling material onto the primary star. From the strongly unstable spiral arms, companions from via fragmentation. After fragmentation, the spin-up event continues until the dense gas has been accreted, or redistributed, and a companion spins up and stabilises the circumstellar disc around the primary star.

For the weakest spin-up event (\Cref{fig:Toomre_Q_S35M0}, we see that many low-mass companions already exist before the spin-up event begins. These companions formed from the ring fragmentation at large separations ($>100\au$), described at the beginning of \Cref{sec:results}. Before the spin-up event, the primary star hosts a circumstellar disc of radius $\sim50\au$, which is marginally unstable ($Q\sim1$). The spin-up event is triggered when the northern companion travels towards the primary, triggering a spiral arm, which is marginally unstable. The triggering of spiral arms due to interactions with a companion or flybys has been documented in previous star formation simulations \citep{kuruwita_dependence_2020, pfalzner_close_2021, cuello_close_2023}. If the angular velocity of the flyby/companion star is lower than the angular velocity of the outer disc, material in the outer disc can be slowed down, and begin migrating inwards through the circumstellar disc, forming a spiral arm. In this scenario, angular momentum is transferred from the orbit to the circumstellar disc, and then onto the primary star \citep{kuruwita_dependence_2020}.

A second spiral arm is also triggered by another companion during the spin-up event, which may contribute to further spin-up of the primary star. The spin-up event ends when one of the companions has migrated inwards, spinning up and stabilising the circumstellar disc around the primary star.

\section{Discussion}
\label{sec:discussion}

This work investigates the relation between binary and multiple star formation and the possible formation of fast rotators. To this end, we carried out simulations of a proto-stellar core collapse with varying initial conditions and observed spin-up events in the primary star to varying degrees.

\subsection{The observed spin-up events in combination with disc lifetimes}

After looking in detail at the strongest and weakest spin-up events, we determine that gravitational instabilities in the disc around the primary star trigger the spin-up event. The origin of these instabilities can be from within the disc itself, as seen in the strong spin-up case (c.f.~Fig.~\ref{fig:Toomre_Q_S20M2}), or due to interactions with a companion, as seen in the weakest spin-up case (c.f.~Fig.~\ref{fig:Toomre_Q_S35M0}). Thus, the discs that would produce the strongest spin-up events would also be prone to disc fragmentation, as seen in our strongest spin-up case. This may then contribute to the observations of many fast rotators having nearby (possibly undetected) companions. These companions may have formed from the strong spin-up event, which led to fragmentation.

While these spin-up events alone may not be sufficient to explain fast rotator formation, when coupled with shortened disc lifetimes, due to disc truncation by companions, this may provide a more compelling scenario for fast rotator formation. As observed in \Cref{fig:sink_spin_spec}, after the spin-up event, in some cases the primary star spin would decrease slowly over an extended period. This is due to the primary accreting low-specific angular momentum material, thus reducing the spin. The effect of this slow spin-down can be reduced by shortened disc lifetimes.

Proto-stars are thought to experience a period of spin-down due to disc-locking soon after formation \citep{konigl_disk_1991}. The presence of a circumstellar disc is necessary for disc-locking, therefore, the lifetime of the disc can also play a role in determining the spin of the star. This is the conclusion that is drawn from the analytical models in \citet{gallet_improved_2013, gallet_improved_2015}, which found that shorter disc lifetimes around fast rotators allowed for more accurate modelling of the long-term evolution ($1000\Myr$ after birth) of fast rotators. Circumstellar discs in binaries tend to be truncated if a companion is nearby, as seen here and in simulations by other groups \citep{artymowicz_dynamics_1994}. Disc truncation by nearby companions is also suggested in observations of weaker millimetre flux, as well as direct imaging \citep{harris_resolved_2012, cox_protoplanetary_2017}. The circumstellar disc truncation is likely to lead to shorter disc lifetimes \citep{kraus_role_2012, kuruwita_multiplicity_2018}, which may play a role in halting the spin-down time that proto-stars in binaries have, allowing the stars to maintain higher rotation rates for longer times. External photo-evaporation can also contribute to shortening disc lifetimes \citep{roquette_influence_2021}, which would be enhanced in the clustered environments where multiple star systems form.

While shorter disc lifetimes are likely to play a role in the observed fast-rotator evolution, the models of  \citet{gallet_improved_2013} set the initial rotation rate of their slow and fast rotators at birth. What sets this initial rotation is still in question, and our work suggests that disc fragmentation and interactions with companions may be a viable mechanism to increase the initial rotation of a star in its early evolution.

\subsection{Resolution study to derive protostellar properties}
\label{ssec:resolution_study}

\begin{figure}
    \centerline{\includegraphics[width=1\linewidth]{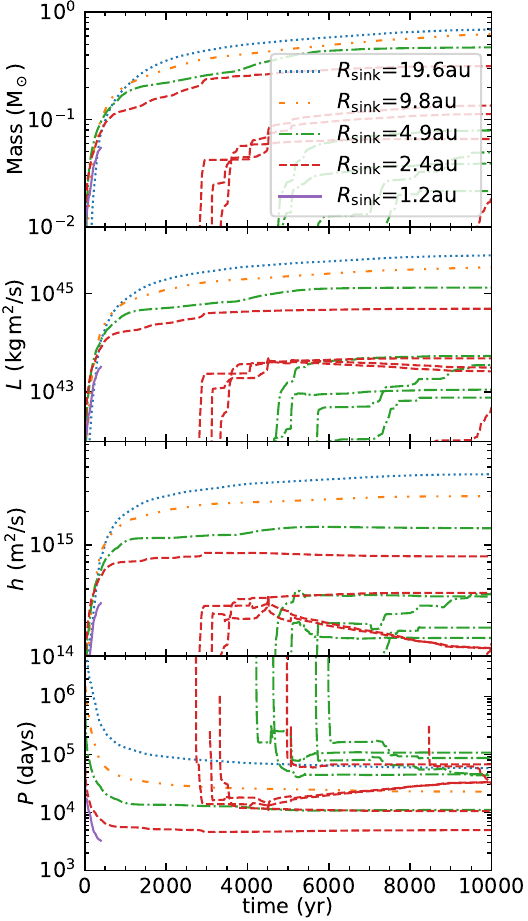}}
    \caption{Time evolution of sink particle mass (\emph{top}), angular momentum (\emph{second panel}), mass-specific angular momentum (\emph{third panel}), and rotation period (\emph{bottom}) of the sink particles, comparing simulations with sink particle radii of $19.6\au$ (dotted blue), $9.8\au$ (dash-dot-dot orange), $4.9\au$ (dash-dotted green), $2.4\au$ (dashed red), and $1.2\au$ (solid purple line).}
    \label{fig:res_study_raw}
    \vspace{-0.5cm}
\end{figure}

\begin{figure}
    \centerline{\includegraphics[width=1\linewidth]{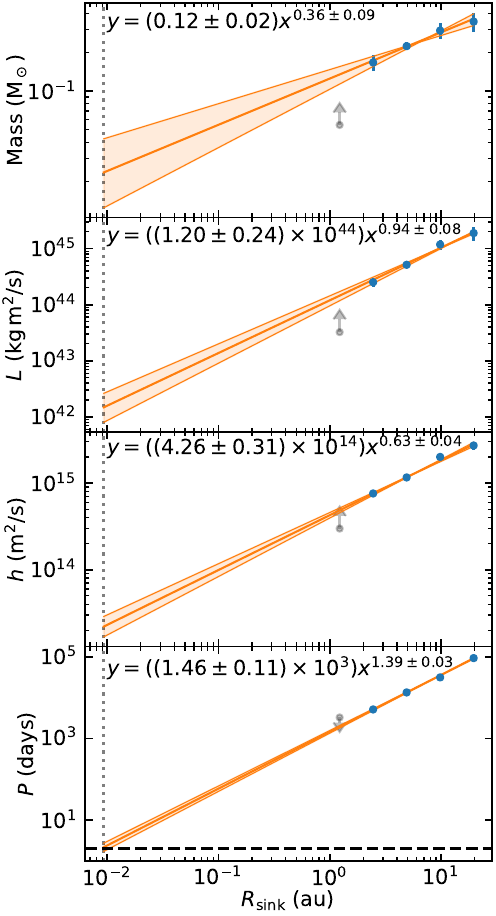}}
    \caption{Resolution study, showing the primary star mass (\emph{top}), angular momentum (\emph{second panel}), mass-specific angular momentum (\emph{third panel}), and rotation period (\emph{bottom}) as a function of $R_\mathrm{sink}$. For each of the blue data points, the error bars are the minimum and maximum values measures between $1500\yr<t<2500\yr$, and the point is the average of these. The orange line shows the derived power-law relationships and the transparent region shows the spread in the fit. The derived fit and error are annotated in the top left of each panel. The grey points show the quantities of the $R_\mathrm{sink}=1.2\au$ simulation, which has not reached $1500\yr$ and is therefore excluded from the fitting (only upper or lower limits are indicated). The vertical dashed line marks $R_\mathrm{sink}=2\rsun\sim10^{-2}\,\au$, and the horizontal dashed line in the bottom panel shows the $2$~day threshold that defines fast rotators \citep{kounkel_measurement_2023}.}
    \label{fig:res_study_scaled}
    \vspace{-0.5cm}
\end{figure}

\begin{figure*} \centerline{\includegraphics[width=0.8\linewidth]{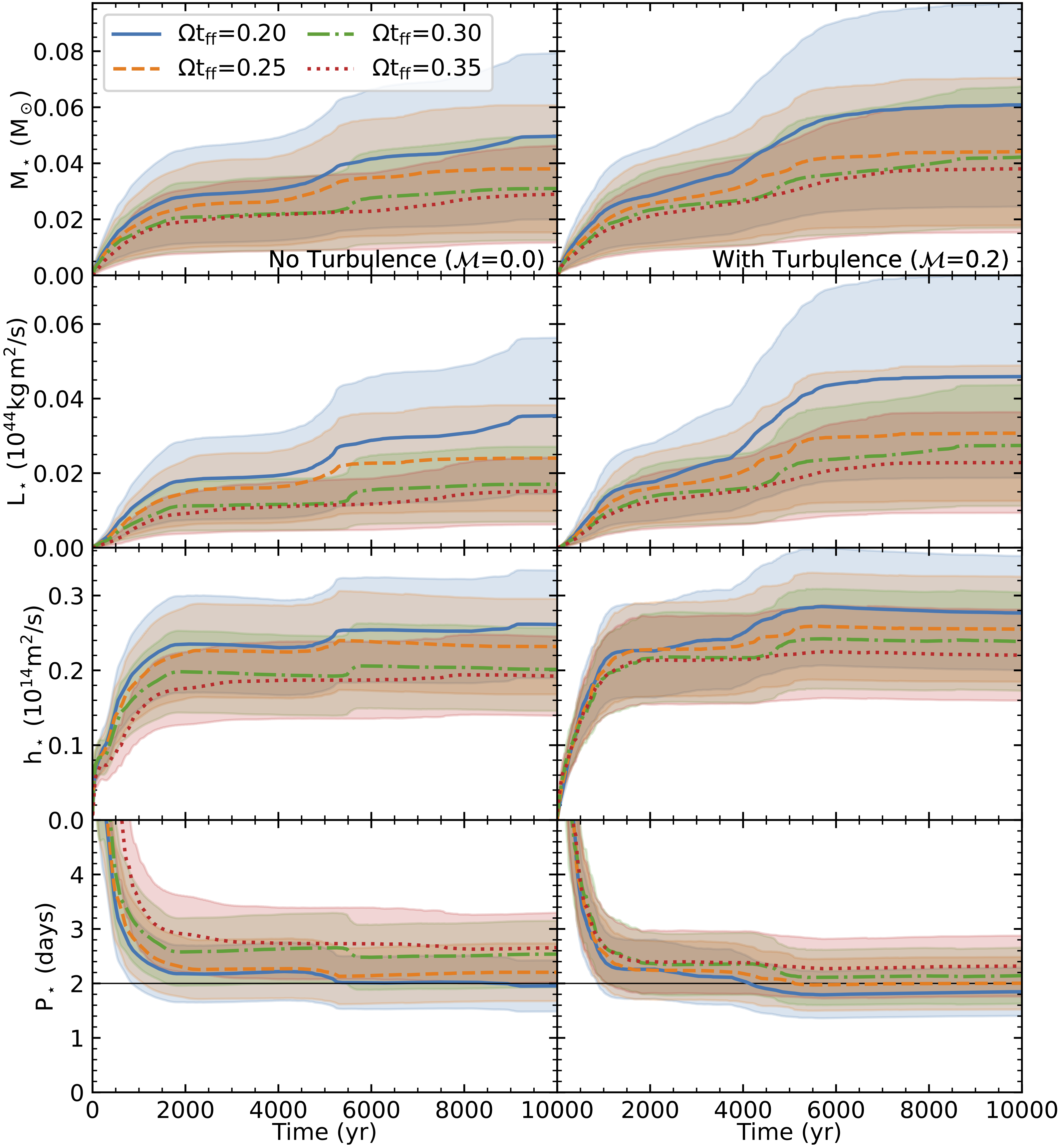}}
    \caption{Derived protostellar mass (\emph{top}), angular momentum (\emph{second row}), specific angular momentum (\emph{third row}), and rotation period (\emph{bottom}), for the non-turbulent (\emph{left}), and turbulent simulations (\emph{right}). The shaded regions are estimated upper and lower bounds of the quantity based on the resolution study, using accretion fractions summarised in \Cref{tab:accretion_fractions}. The solid lines are the means of the upper and lower bounds. The horizontal black line at 2~days highlights the boundary between fast rotators and non-fast rotators \citep{kounkel_measurement_2023}.}
    \label{fig:protostar}
\end{figure*}

Throughout this work, we have used a sink particle method to study how proto-stars accrete angular momentum. While we have frequently referred to the sink particles as stars, the particle is actually modelling a combination of the star and inner disc, because it has an accretion radius of $4.9\au$. It is currently not possible to run these simulations for a long evolutionary time and resolve the stellar surface at the same time, due to the prohibitively high computational cost. We instead ran a resolution study on the simulation that showed the strongest spin-up event, to understand the dependence of the sink particle properties on resolution.

In this resolution study, we compare simulations with sink particle accretion radii of R$_\mathrm{sink} = 19.6, 9.8, 4.9, 2.4, 1.2\au$. The resulting accreted mass, total angular momentum ($L$), and specific angular momentum ($h$) of the sink particles are shown in \Cref{fig:res_study_raw}. The rotation period of the sink particle is calculated using \Cref{eqn:rotation_rate}.

In all the panels of \Cref{fig:res_study_raw}, we see that the mass, momentum and specific angular momentum of the primary star decrease with increasing resolution. This is because, as the accretion radius decreases, more mass remains in the gas near the sink particle in the inner disc, rather than being accreted onto the sink particle. With higher-resolution simulations, we are also better resolving the high-velocity jet and wind-launching region near the proto-star. The jets carry away mass and angular momentum from the circumstellar disc, reducing the mass and angular momentum accretion rate onto the star, as seen in the simulations before the formation of companions at around $2500\yr$.

The angular momentum that is accreted onto a sink particle is directly proportional to the sink particle radius because angular momentum is proportional to $\mathbf{r} \times \mathbf{v}$. The maximum radius that a sink particle can accrete from is defined by the accretion radius, therefore, a larger sink particle radius means more accreted angular momentum.

The resulting rotation period is directly proportional to the sink particle accretion radius and specific angular momentum as demonstrated in \Cref{eqn:rotation_rate}, showing $P_\mathrm{sink}\propto r_\mathrm{sink}^2/h_\mathrm{sink}$. Thus, with higher resolution (lower $r_\mathrm{sink}$), the rotation period of the sink particle decreases. The rotation periods that we can directly resolve with these simulations are orders of magnitude larger than the $2\days$ threshold that defines fast rotators \citep{kounkel_measurement_2023}.

\begin{table}
	\centering
	\begin{tabular}{|l|c|c|c|c|}
		\hline
		  Quantity  & lower efficiency & higher efficiency\\
    \hline
          $M$ (M$_\odot$) & 0.0521 & 0.2065\\
          $L$ (kg$\,$m$^2$/s) & 0.0014 & 0.0056\\
          $h$ (m$^2$/s) & 0.0143 & 0.0251\\
          $P$ (days) & 0.0001 & 0.0002\\
		\hline
	\end{tabular}
	\caption{Fraction of sink particle quantity at $r_\mathrm{sink}=4.9\au$ expected to be in a protostar of radius $r=2\rsun$ derived from the power law fits shown in \Cref{fig:res_study_scaled}.}
	\label{tab:accretion_fractions}
\end{table}

Based on the trends found in this resolution study, we derive relationships between the measured quantities and the resolution (parameterised by $R_\mathrm{sink}$). To do this, we show the primary star's mass, momentum, specific angular momentum, and rotation period, against the sink particle accretion radius in Fig.~\ref{fig:res_study_scaled}. We measure the primary star properties by taking the average of the minimum and maximum values between $1500\yr<t<2500\yr$ (c.f., Fig.~\ref{fig:res_study_raw}). The period of $1500\yr<t<2500\yr$ was chosen because the quantities are beginning to plateau in this time window, and companions have not formed yet, as we want to avoid companion formation to interfere with a clean resolution study. The primary star values are plotted in blue in \Cref{fig:res_study_scaled}. The final values of the $R_\mathrm{sink}=1.2\au$ simulation are also plotted in grey as lower and upper limits (depending on the quantity in each panel), even though it has not reached $1500\yr$. The intervals between the quantities in all the other simulations ($R_\mathrm{sink}>1.2\au$) are relatively even, suggesting a power-law relationship between the quantity and the resolution. We use the \texttt{curvefit} function in the python package \texttt{scipy}, to derive power-law fits to quantify the relationship, which are shown in orange in \Cref{fig:res_study_scaled} and annotated in the top left of each panel. The values from the $R_\mathrm{sink}=1.2\au$ were not included in the fitting.

We extend the derived power laws down to resolutions where the sink particle radius would be $2\rsun$, which is annotated by the vertical grey dotted line. We see that at $R_\mathrm{sink}=2\,\rsun\sim10^{-2}\,\au$, the rotation period of the sink particle would be $<10\days$, which is consistent with observations. The dashed line in the bottom panel of \Cref{fig:res_study_scaled} indicates the $2$~day threshold that defines fast rotators, and we see that the extrapolation based on our resolution study reaches this limit. The comparison is taken before companion formation, which we have shown triggers spin-up events within the primary star, therefore, at resolutions where the stellar surface is resolved and there is companion formation, the rotation period is expected to be even shorter.

Based on the extrapolations shown in \Cref{fig:res_study_scaled}, we calculate what fraction of a sink particle quantity would reside in a proto-star of radius $r=2\rsun$ for all simulations. The lower and upper limits of these fractions are summarised in \Cref{tab:accretion_fractions}. We use these fractions to derive the properties of the primary star in our simulations. We see that due to the large uncertainties in the accretion efficiencies, there is significant spread and overlap in the derived periods of the primary star, see in the bottom panel of \Cref{fig:protostar}. Despite the uncertainties, we see that the spread of periods of the simulations with stronger spin-up events ($\Omega t_\mathrm{ff}=0.20$) are smaller with a low mean period, compared to the simulations with the weakest spin-up events ($\Omega t_\mathrm{ff}=0.35$). In the bottom panel of \Cref{fig:protostar}, we mark the $2$~day threshold used to define fast rotators, and we see that the mean rotation period of our simulation with the strongest spin-up passes this threshold after its spin-up event.

\section{Limitations and caveats}
\label{sec:caveats}

\subsection{Numerical resolution}

As discussed in \Cref{ssec:resolution_study}, the chosen resolution in our simulations does not resolve the inner circumstellar disc where jets and high-velocity outflows are launched. These mechanisms are important for regulating the mass and angular momentum that is accreted onto the protostar.

To understand the effect of resolution on the initial spin-up of stars, we carried out a resolution study on the turbulent $\Omega t_\mathrm{ff}=0.20$ simulation, which is presented in \Cref{ssec:resolution_study}. In this study, we find that the formation of companions happens earlier in higher-resolution simulations than in lower-resolution simulations. This is likely because the small-scale turbulent modes are better resolved with higher resolution, and if fragmentation starts here, then the instability can grow faster if better resolved early on.

From the resolution study, we also observe that the rotation period of the sink particles measured after their initial spin-up during the protostellar collapse, but before the formation of companions, decreases with resolution. In Fig.~\ref{fig:res_study_scaled}, we extrapolated what the sink particle period would be at resolutions where the protostellar surface is resolved (R$_{sink}\sim2\rsun$; \citet{federrath_modeling_2014}), and determined that this is well within observed rotation rates ($<10\days$).

Another effect of resolution and finite numerical precision is that asymmetries can be introduced into symmetric setups. The simulation with $\Omega t_\mathrm{ff}=0.20$ and no turbulence was intended to be a single-star simulation. This is a symmetric setup that should produce a single star as seen in previous works \citep{federrath_modeling_2014, kuruwita_binary_2017, gerrard_role_2019}, however, those works only ran the simulations for at most $3000\yr$. In this work, we have pushed this simulation to $10\,000\yr$. A result of this is that, despite the initially symmetric setup, due to finite numerical precision, asymmetries are introduced that are amplified. This then leads to fragmentation later in the simulations.

However, we argue that before fragmentation, the systems are appropriate approximations of a single-star system. This is why we compare the spin of the primary star before and after companion formation. In most simulations, the primary star's evolution in mass, angular momentum, and specific angular momentum is relatively steady before companion formation. Thus, as long as the companion(s) has/have not formed yet, the primary star's evolution does not show evidence of spin-up.

\subsection{Grid refinement}

The initial conditions for the non-turbulent simulations are inherently symmetric, however, asymmetries can grow due to numerical round-off errors. Some of these round-off errors can be introduced where the AMR grid refines. In the case of the high-spin, non-turbulent simulations, the formation of a ring is physical, because it is a natural consequence of the high initial angular momentum of the protostellar cloud. However, the ring fragmentation is seeded by the grid noise, which is aligned with the grid. In this sense, the non-turbulence simulations are less reliable in what happens concerning fragmentation. However, despite the numerical seed of fragmentation, further asymmetries develop, and the system `forgets' its initial conditions and evolves more realistically. The impact of the grid refinement on fragmentation is not seen in the turbulent simulations, as the fragmentation is seeded by physical perturbations in that case.

\subsection{Radiation feedback}
\label{ssec:radiation}

Radiation feedback is important for various aspects of star formation including heating gas to stabilise it against collapse or fragmentation, and changing ionisation fractions which sets the coupling between gas and magnetic fields \citep{wunsch_radiation_2024}. Our simulations do not explicitly calculate radiative transfer, however, our equation of state accounts for some of the radiative effects on the local cell scale (see \Cref{sec:method}).

Most numerical work simulating radiation feedback has focused on massive star formation \citep{cunningham_radiation-hydrodynamic_2011, harries_radiation-hydrodynamical_2017, kuiper_first_2018, mignon-risse_disk_2023}. The consensus is that radiative feedback tends to reduce the formation of higher-order multiples, by reducing providing thermal support against fragmentation during massive star formation \citep{offner_investigations_2014, rosen_massive-star_2019}.

Radiation hydrodynamic simulations of low-mass protostellar collapse find that the initial collapse into a protostar is prolonged by radiative thermal support \citep{ahmad_birth_2023}. However, if the protostellar core is massive enough, the inward pressure of the gravitational collapse can easily overwhelm the outward radiation pressure \citep{vaytet_grid_2017, bhandare_first_2018}, 

After the initial collapse of a protostellar core into a protostar, radiation feedback from the central star on the circumstellar disc will stabilise the disc against fragmentation \citep{bate_stellar_2012}. Hydrodynamic simulations including radiation feedback have found that the primary star mass is higher because fragments are not forming from the over-densities and are instead being accreted onto the primary star, \citep{offner_investigations_2014}. Accounting for thermal support in the creation of sink particles from over-densities in a circumstellar disc may be important to prevent artificial fragmentation. This is because over-densities that form can later be destroyed by shear turbulent motions within the disc \citep{oliva_modeling_2020}. \citet{bate_stellar_2012} conclude that the main physical processes involved in determining the properties of multiple stellar systems are gravity and gas dynamics.

Radiation smoothed particle hydrodynamic simulations of stellar flybys and triggered sub-stellar companion formation work carried out by \citet{cadman_binary_2022}. The authors find that flybys at separations of $100<r<400\au$ could make a marginally stable circumstellar disc fragment, which is a similar conclusion to our results. The simulations of \citet{cadman_binary_2022}, however, neglect magnetic fields. The inclusion of both radiation and magnetism would likely lead to stronger support against gravitational collapse, limiting fragmentation.

Works combining radiation and magnetohydrodynamics (RMHD) looking at protostellar collapse have been carried out, but often struggle to progress beyond the first $2000\yr$ after protostar formation. RMHD simulations find that the radius of the first protostellar core is larger with radiation \citep{tomida_radiation_2010}. RMHD simulations find that magnetism and thermal forces both contribute to the launching of outflows \citep{bate_collapse_2014}, affecting the efficiency of mass and momentum lost in outflows. Simulations with radiation and non-ideal MHD will be discussed in \Cref{ssec:non_imhd}.

The inclusion of radiative feedback in our simulations would reduce the amount of fragmentation observed. Stronger gravitational instabilities would be required for the simulated circumstellar discs to fragment. Our simulation with the largest spin-up event showed a strong gravitationally unstable spiral arm formed that fragmented. With radiation, the instability may last longer, leading to a longer and stronger spin-up event before fragmentation is triggered.

\subsection{Non-ideal MHD effects and resolving outflows}
\label{ssec:non_imhd}

Our simulations use ideal MHD, however it is known that various non-ideal MHD effects begin to affect evolution on disc scales. The non-ideal effects of Ohmic resistivity, the Hall effect and ambipolar diffusion are important on scales $\sim$1.5, $2$--$3$ and $\geq 3$ scale heights, respectively \citep{wardle_magnetic_2007, salmeron_magnetorotational_2008, konigl_effects_2011, tomida_radiation_2015, marchand_chemical_2016}. Further away from the disc, the surface layers of discs are expected to be ionised by stellar radiation in the FUV and the ideal MHD limit is a reasonable approximation \citep{perez-becker_surface_2011, nolan_centrifugally_2017}.

A common problem with ideal MHD has been that angular momentum is too efficiently removed from collapsing protostellar cores, suppressing the formation of large circumstellar discs. This has been called the ``magnetic braking catastrophe''. Non-ideal MHD provides a solution to this by changing the strength of coupling between the magnetic fields and gas based on ionisation fractions \citep{wurster_can_2016}.

Simulations with ambipolar diffusion find that magnetically driven outflows are weakened due to the diffusion \citep{masson_ambipolar_2016}. While our simulations do no explicitly calculate ambipolar diffusion, there is numerical diffusion, which has been argued to reproduce the effect of ambipolar diffusion at large scale-heights \citep{hennebelle_role_2019}.

Simulations with Ohmic resistivity, which is most important in the mid-plane of circumstellar discs, find that because angular momentum is not efficiently removed, the disc surface density increases and gravitational instabilities form \citep{machida_first_2019}.

Many simulations combining ambipolar diffusion and Ohmic resistivity continue to find the angular momentum removal via jets and outflows are weakened, helping in the creation of larger circumstellar discs \citep{dapp_bridging_2012, vaytet_protostellar_2018, higuchi_driving_2019, marchand_protostellar_2020}. Simulations of the Hall effect confirm that the orientation of the rotation and magnetic field influences the size of discs formed \citep{zhao_interplay_2021}.

When all three non-ideal effects are simulated, it is difficult to determine which effects dominate in various scenarios. \citep{wurster_impact_2021} argues that the Hall effect has the greatest influence on disc formation and outflow efficiencies, but other work finds that dust distributions can influence whether the magnetic couple of ambipolar diffusion or the Hall effect dominates gas dynamics \citep{marchand_protostellar_2020, zhao_interplay_2021}. While simulations of Ohmic diffusion find that gravitationally unstable discs form more easily, the inclusion of the other non-ideal effects may stabilise discs \citep{tu_protostellar_2024}. 

The overall conclusion is that non-ideal MHD effects reduce angular momentum removal efficiencies via jets and outflows, and help with the formation of large circumstellar discs. Because larger more massive discs are formed with non-ideal MHD, this increases the number of binary and multiple stars formed via disc fragmentation \citep{wurster_impact_2017}. Therefore, the inclusion of non-ideal effects in our simulation may aid fragmentation, which is at odds with the impact of radiation feedback, which we concluded at the end of \Cref{ssec:radiation}.

Non-ideal MHD effects are sensitive to the ionisation fraction of the gas, and radiation plays a role in setting these fractions, therefore, combining non-ideal MHD and radiative feedback is important to building a complete physical model of protostellar core collapse. Simulations aiming to combine some non-ideal effects and radiation find that while radiation pressure can contribute to outflows, jets and outflows are dominated by magnetism \citep{commercon_discs_2022}. The degree of ionisation influences outflow speeds \citep{wurster_collapse_2018}, but overall non-ideal effects reduce angular momentum removal \citep{tomida_radiation_2015}.

The inclusion of non-ideal effects in our simulations would weaken the angular momentum loss via outflows, leading to higher angular momentum being accreted onto the sink particles. However, due to resolution limits, we only resolve the low-velocity components, and we do not expect the observed sink-particle spin evolution to differ significantly with non-ideal MHD. The calculated protostellar rotation periods are significantly more sensitive to the assumption of mass and momentum accretion from the inner disc onto the protostar.

\section{Summary and Conclusion}
\label{sec:conclusion}

In this paper, we investigated whether the formation of binary and multiple-star systems inherently leads to the formation of fast-rotating stars. This is inspired by observations suggesting that many fast rotators are expected to have a nearby, low-mass, unseen companion. We use MHD simulations of the collapse and fragmentation of turbulent and non-turbulent protostellar cores to shed light on this subject. We use a sink particle method to study the spin evolution of the stars that form. We find that in all simulations there is significant spin-up in the primary star, which is correlated with the formation of companions (\Cref{fig:sink_spin_spec}). We further find that turbulent simulations produce stronger spin-up events than simulations without turbulence (\Cref{fig:spin_up}). Upon further inspection, we identified two mechanisms that lead to these spin-up events:

\emph{Strong gravitational instability in disc leading to fragmentation}: In the simulation with the strongest observed spin-up event ($18\%$ increase in specific angular momentum, as quantified in Fig.~\ref{fig:spin_up}) we find that the event begins before the formation of companions. The spin-up event is triggered by a strongly unstable perturbation (as quantified by Toomre~$Q$) in the circumstellar disc, which creates spiral arms that funnel a large amount of specific angular momentum onto the primary star. Companions are eventually formed by fragmentation of the unstable spiral arms within the disc, which slows down the spin-up event (\Cref{fig:Toomre_Q_S20M2}).

\emph{Stellar flybys triggering a weaker gravitational instability}: In the simulation with the weakest observed spin-up event ($5\%$ increase in specific angular momentum, as quantified in Fig.~\ref{fig:spin_up}), we find that many companions always form at large separations. The spin-up event is triggered when one of these companions flies by the primary star, perturbing the marginally Toomre-$Q$ stable circumstellar disc, and triggering the formation of a spiral arm, which spins up the primary star. Further fragmentation is not triggered in this scenario, however, the flyby companion settles into a lower separation orbit (\Cref{fig:Toomre_Q_S35M0}).

\emph{Protostellar properties}: We derive protostellar parameters from the sink particle data of the primary stars based on a resolution study (\Cref{fig:res_study_raw}). We find that the simulations with the strongest spin-up events have lower period ranges, entering the fast rotator regime (\Cref{fig:res_study_scaled}).

Overall, we conclude that there is a causal relationship between strong spin-up events and the formation of nearby companions via disc fragmentation. The formation of nearby companions can also lead to truncated circum-stellar discs and shorter spin-down times. The combination of strong spin-up events creating companions and disc truncation by these companions can explain why many fast rotators have unresolved nearby low-mass companions.

\section*{Acknowledgements}

RLK acknowledges funding from the Klaus Tschira Foundation. C.F.~acknowledges funding by the Australian Research Council (Discovery Projects grant~DP230102280), and the Australia-Germany Joint Research Cooperation Scheme (UA-DAAD). C.F.~further acknowledges high-performance computing resources provided by the Leibniz Rechenzentrum and the Gauss Centre for Supercomputing (grants~pr32lo, pr48pi and GCS Large-scale project~10391), the Australian National Computational Infrastructure (grant~ek9) and the Pawsey Supercomputing Centre (project~pawsey0810) in the framework of the National Computational Merit Allocation Scheme and the ANU Merit Allocation Scheme.
yt \citep{turk_yt:_2011} was used to help visualise and analyse these simulations. The simulation software, \texttt{FLASH}, was in part developed by the Flash Centre for Computational Science at the University of Chicago and the Department of Physics and Astronomy at the University of Rochester.

\bibliographystyle{aa}
\bibliography{references_clean.bib}

\end{document}